\documentstyle[12pt,epsfig]{article}

\newskip\humongous \humongous=0pt plus 1000pt minus 1000pt
  \newif\ifdtup

\def\frac#1#2{ {{#1} \over {#2} }}

\def\ie{\hbox{\em i.e. }}

\def\beq{\begin{equation}}
\def\eeq{\end{equation}}

\def\non{\nonumber}
\def\beqn{\begin{eqnarray}}
\def\eeqn{\end{eqnarray}}

\def\L{\Lambda}
\def\be{\beta}

\def\de{\delta}

\def\g{\gamma}
\def\as{\alpha_{\sf s}}
\def\a0{\alpha_{\sf 0}}
\def\aV{\alpha_{\sf V}}

\def\dm{\delta m}

\def\MSbar{\overline{\rm MS}}

\def\Journal#1#2#3#4{{#1} {\bf #2}, #3 (#4)}
\def\NPB{{\em Nucl. Phys.} B}
\def\NPBPS{{\em Nucl. Phys.} B (Proc. Suppl.)}
\def\PLB{{\em Phys. Lett.}  B}
\def\JHEP{{\em Journ. of High Energy Physics}}
\def\PRL{\em Phys. Rev. Lett.}
\def\PR{\em Physics Reports}

\begin{document}
\begin{titlepage}
\begin{flushright}
     UPRF-2000-18 \\
     SWAT:282 \\
     December 2000\\
\end{flushright}
\par \vskip 10mm
\begin{center}
{\Large \bf
The residual mass in Lattice Heavy Quark Effective Theory to $\alpha^3$ 
order \footnote{Research supported 
in part by Italian MURST under contract 9902C68583, by I.N.F.N. 
under i.s. PR11 and by EU Grant EBR-FMRX-CT97-0122.}}
\end{center}
\par \vskip 2mm
\begin{center}
F.\ Di Renzo$\,^a$,
and  L.\ Scorzato$\,^{a,b}$ \\
\vskip 5 mm
$^a\,${\it Dipartimento di Fisica, Universit\`a di Parma \\
and INFN, Gruppo Collegato di Parma, Italy}\\
\vskip 2 mm
$^b\,${\it Department of Physics, \\
University of Wales, Swansea, U.K.}
\end{center}
\par \vskip 2mm
\begin{center} {\large \bf Abstract} \end{center}
\begin{quote}
We determine to order $\alpha^3$ in the quenched approximation the 
so--called residual mass in the lattice regularisation of the Heavy 
Quark Effective Theory. We follow a gauge--invariant strategy 
which exploits the fact that this mass term dominates the exponential 
decrease of perturbative Wilson loops at large perimeters. Our 
computational tool is Numerical Stochastic Perturbation Theory. 
The new coefficient we compute is crucial to improve the determination 
of the ($\MSbar$) mass of the $b$--quark from lattice simulations of 
the Heavy Quark Effective Theory.
\end{quote}

\end{titlepage}

\section{Introduction}

Heavy Quark Effective Theory (HQET) is proving to be a particularly 
useful tool for phenomenological studies in heavy flavour physics 
(see \cite{HQET} for a review and references therein to original papers). 
In this approach physical quantities are expanded as series
in inverse powers of the heavy quark masses $m_Q>>\L_{QCD}$. 
The Lagrangian density of the effective theory may be written as
\beq \label{HQETaction}
{\cal L}_{HQET} = \overline{h} D_4 h,
\eeq
where $D_4$ is the (gauge) covariant time derivative; $h$ is
the bi--spinor describing the only degrees of freedom of the heavy quark
that are relevant in this approximation.
Soon after its introduction it was realized that while HQET 
could appear conceptually simple, there was a variety of theoretical 
subtleties one should face in order to profit from its application. 
Not surprisingly, many of these subtleties have to do with the definition 
of the heavy quark mass $m_Q$ itself. First of all it was realized 
\cite{FNL92} that the theory must be invariant by a redefinition 
of $m_Q$ which leaves the momentum $P^\mu_Q$ fixed:
\beq
P^\mu_Q = m_Q v^\mu + k^\mu.
\eeq
In fact the mass of a quark is not a well defined physical quantity, and
the separation of $P^\mu_Q$ into a large and a small ($k^\mu \sim \L_{QCD}$)
component is to some extent arbitrary. The requirement of invariance by
redefinition of $m_Q$ leads to the introduction of a ``covariant''
derivative in terms of which the HQET Lagrangian can acquire a mass term 
for the static quark. It was noted however that even so the expansion
in $1/m_Q$ is not well defined and other ambiguities show up. In dimensional 
regularisation they appear as UV renormalons in the matrix elements of
HQET and IR renormalons in the coefficient functions that match to QCD
\cite{BB94}. In the lattice regularisation problems show up
as non--perturbative power divergences in the inverse lattice spacing
\cite{MMS92}. The ambiguities in the coefficient functions are expected
to cancel with those in the HQET matrix elements. In order to perform 
such a subtraction and to obtain a properly defined expansion the expansion 
parameter itself ($m_Q$), first of all,  must
be defined in a way free of ambiguities. The pole mass $m_{pole}$
proved not to be a good definition \cite{BB94}. The authors of
\cite{MS95} introduced a lattice non--perturbative definition for the
quark mass (the so--called subtracted pole mass) which is free of 
renormalon ambiguities and power divergences 
\beq\label{subpolemass}
m^S_Q \equiv m_H - \overline{\L}. 
\eeq
In the previous formula $m_H$ is the mass of an hadron containing the 
heavy quark, and
$\overline{\L}$ is the non--perturbative definition of a subtracted 
binding energy
\beq\label{subinden}
\overline{\L} \equiv {\cal E} - \de\overline{m}.
\eeq
${\cal E}$ is the ``naive'' binding energy, computed as the decay constant 
of the correlation function of two axial currents 
($A_4=\overline{h(x)}\g_4\g_5 q(x)$, $q(x)$ being a light quark) evaluated 
with the action in Eq.~(\ref{HQETaction}):
\beqn
C(t) &=& \sum_{\vec{x}} \langle A_4(\vec{x},t)A_4(\vec{0},0)^\dag \rangle \\
&\sim & Z^2 \exp(-{\cal E}t).
\eeqn
The other term in Eq.~(\ref{subinden}) is a non--perturbative definition of 
the so--called residual mass counterterm. The latter is just the quantity 
we are concerned with in this paper. It is a linearly divergent mass 
counterterm which quantum corrections generate in the lattice regularisation, 
contrary to dimensional regularisation. In the non perturbative approach we 
have just referred to, it is fixed by the large time behaviour of the quark 
propagator $S_h^{ij}(x,y)$ ($i,j$ colour indexes):
\beq
\de\overline{m}\equiv
- \lim_{t\rightarrow\infty} \frac{1}{a} 
\log\left(\frac{|S_h^{ii}(\vec{x},t+a;0,0)|}{|S_h^{ii}(\vec{x},t;0,0)|}\right).
\eeq
Both $m^S_Q$ and $\overline{\L}$ are non perturbative, well defined 
quantities and may be evaluated by numerical simulations on the lattice 
\cite{CGMS95,GGMR00}. This is not at all a trivial observation. 
Both ${\cal E}$ and $\de\overline{m}$ are linearly divergent quantities on 
their own and only the result of the subtraction in Eq.~(\ref{subinden}) is 
a well defined, divergences free quantity, which also makes well defined 
Eq.~(\ref{subpolemass}). 
However, the subtracted pole mass is still a long distance quantity and in 
order to be able to extract physical predictions from lattice HQET 
it is necessary to match the definition of $m^S_Q$ to another mass 
definition such as the mass $\overline{m_Q}$ of $\MSbar$. In fact the 
coefficient functions of the
$1/m_Q$ expansion are typically computed in perturbation theory in 
dimensional regularisation \cite{MS95}. 
The relation between $m^S_Q$ and the $\MSbar$ mass of the $b$--quark
$\overline{m_b}(\overline{m_b})$ has been computed in \cite{CGMS95,GGMR00} 
and exploited to compute the mass of the $b$--quark. 
The strategy goes as follows. The pole mass 
(we write $m_b^{pole}$ having in mind the pole mass for the $b$--quark) 
can be related to the $\MSbar$ mass in perturbation theory
\beq \label{matching0}
\overline{m_b}(\overline{m_b}) = 
m_b^{pole} \, 
\left [ 1 + \sum_{n=0}^{\infty} (\frac{\as(\overline{m_b})}{\pi})^{n+1}
D_n \right ]
\eeq
Note that in this relation coefficients $D_n$ with $n \leq 2$ are known 
\cite{CSR99,MR99}. By matching the propagator in QCD to its lattice 
HQET counterpart (we closely follow \cite{GGMR00}) one can then relate the 
pole mass to the binding energy and the physical mass of an hadron, for 
example the B meson mass ($M_B$). 
This relation is just the perturbative version of Eq.~(\ref{subpolemass})
\beq
m_b^{pole} = M_B - {\cal E} + \sum_{n=0}^{\infty} \as^{n+1} \frac{X_n}{a} 
+ {\cal O}(1/m_b)  
\equiv M_B - {\cal E} + \dm + {\cal O}(1/m_b)  
\eeq
This time the residual mass term ($\dm$) has naturally emerged from the 
perturbative computation of the heavy quark propagator on the lattice. 
We slightly change notation to stress that $\dm$ is the perturbative 
residual mass. By substitution one gets 
\beq \label{matching}
\overline{m_b}(\overline{m_b}) = 
\left [ M_B - {\cal E} + \sum_{n=0}^{\infty} (\as(\overline{m_b}))^{n+1}
\frac{X_n}{a}\right ]
\left [ 1 + \sum_{n=0}^{\infty} (\frac{\as(\overline{m_b})}{\pi})^{n+1}
D_n \right ]
\eeq
Note that the last expression asks for both a non--perturbative (${\cal E}$) 
and a perturbative ($\dm$) computation. $\dm$ is still in charge of 
cancelling the linear divergence of ${\cal E}$, even if now the 
cancellation is in perturbation theory, so that it is incomplete. What 
one also gets from Eq.~(\ref{matching}) is a delicate cancellation of 
renormalon ambiguities as well: the renormalon in the expansion of $\dm$ 
cancels the one in the perturbative relation between the pole and $\MSbar$ 
masses and this is the key issue. This cancellation asks first of all for 
the use of the same coupling in both perturbative expansions. It also asks 
for high orders in the perturbative expansions. Since, as we have already 
said, $D_2$ is known, this means that $X_2$ has to be computed. In the 
end it turns out \cite{GGMR00} that the ignorance of this term is 
the main source of error in the final determination of 
$\overline{m_b}(\overline{m_b})$. This observation ends this 
introductory review which sets the motivations for our computation: 
the purpose of this paper is to compute $\dm$ to $\as^3$ order ($X_2$) 
in the quenched approximation. Our computational tool is Numerical 
Stochastic Perturbation Theory (NSPT) \cite{NSPT}. We point out that 
while $X_0$ has been known for a long time, $X_1$ has only recently 
been computed \cite{MS98}, while the only results available for $X_2$ 
are at the moment the ones presented at the Lattice 99 conference 
\cite{lat99,mack}, which we will compare our result to. 
We stress that the present computation is still 
a partial result, since we stay within the quenched approximation. 
We emphasize that an unquenched, non--perturbative computation of the 
binding energy is already available, so that an extension of our result to 
include the fermionic loops contribution is compelling. We are 
actually working on it. With this respect the present computation is a 
first step which nevertheless we think is useful for at least a couple of 
reasons. First of all, we will show that NSPT can get the result to an 
accuracy which is good enough for the purpose at hand. On top of that we will 
discuss how our result has an impact on the final accuracy on 
$\overline{m_b}(\overline{m_b})$, at least in the quenched approximation. 
In the next section we briefly summarize the strategy of our computation 
and discuss the numerical analysis which needs to be performed. In section 
3. we report our results and their impact. Finally in section 4. we 
state our conclusions. 

\section{The computational strategy}
An obvious choice in order to compute the residual mass would be the 
computation of the heavy quark propagator. This would ask for gauge 
fixing. While our computational tool (NSPT) can manage such a 
task, it turns out that gauge invariant computations are in NSPT 
both easier and less time demanding (for further technical details on 
this and other points concerning the method itself, see \cite{NSPT}). 
The fact that a gauge invariant computation can be set up in order to obtain 
the residual mass has already been pointed out and exploited in 
\cite{MS98} and also in \cite{lat99,mack}. In order to understand, 
consider a gauge invariant loop such as a Wilson ($W$) or a Polyakov ($P$) 
loop. There are three sources of divergences for such a quantity \cite{DV80}. 
A first one is the logarithmic divergence that can be as usual absorbed 
in the definition of the renormalised coupling. An additional logarithmic 
divergence appears if the contour has an angle (\ie such a problem is 
there in the case of a Wilson loop and not for a Wilson line or a Polyakov 
loop); this is 
usually referred to as the corner divergence. A third, linear divergence 
is the one we are interested in, resulting in (consider for example a 
Wilson loop and note that the divergence exponentiates and shows up as 
a factor) 
\beq\label{Wlindiv}
\langle W \rangle = \exp({- c \, L/a}) \; W_{log} 
\eeq
In the above formula $L$ is the length of the loop and $a$ the cutoff scale 
at small distance (we have of course already in mind the lattice spacing). 
The notation $W_{log}$ reminds that only logarithmic divergences are left. 
The Wilson line (also called P--line) one would need to compute in order 
to obtain the lattice heavy quark propagator would of course show the 
same divergences. 
It is interesting to note that HQET was not there in the early eighties 
when people first got interested in the so--called "phase factors" in Gauge 
Theories. Still of course the physical picture was already there, the linearly 
divergent factor being regarded as mass renormalisation of a test particle 
if one treats the loop as an effective amplitude for a test particle moving 
along the trajectory given by the loop itself. 
Martinelli and Sachrajda \cite{MS98} exploited the 
above considerations in order to compute $\dm$ to second order by extracting 
the linearly divergent contributions from the whole set of integrals 
contributing to the Feynmann graphs for Wilson loops $W(R,T)$ of generic 
size $R \times T$ in lattice perturbation theory. 
The latter ones had already been computed to order $\as^2$ in \cite{HK}. 
In \cite{lat99} our group pinned down a first number for $\dm$ to third order 
(basically its order of magnitude) by performing a fit of squared Wilson 
loops of different sizes $L$ to the expression given in Eq.~(\ref{Wlindiv}). 
At the same Lattice 99 conference the Fermilab group presented a value for 
the same quantity by considering a Polyakov loop $P$ in which the mass term 
contribution was isolated by the Coulomb (self--)interaction of the heavy 
quark \cite{mack}. Let us now introduce the method we exploited in this work. 
We think that a very neat way to extract $\dm$ goes along the lines we now 
proceed to describe. Consider the quantity 
\beq\label{Vr}
V(R) \equiv \lim_{T\rightarrow\infty} V_T(R); \;\;\;\;\;\;\;\;\;\;\;\;
V_T(R) \equiv \log \left( \frac{W(R,T-1)}{W(R,T)} \right).
\eeq
$W(R,T)$ are Wilson loops of size $R \times T$. 
Note the dependence on $T$ which is removed by taking the $T \rightarrow 
\infty$ limit; we do not further comment on this point at the moment 
(there will be a lot to say about that later). 
Eq.~(\ref{Vr}) goes of course back to the old subject of Creutz's ratios 
defined to extract the static inter--quark potential. It is for example 
considered in \cite{HK}. It actually achieves the separation 
of the different (divergent) contributions we are interested in. 
First of all, note that the corner divergences are cancelled between 
numerator and denominator, the number of corners being the same. 
We now compute everything in lattice perturbation theory. In the end one 
is left with 
\beq\label{Vreq}
V(R) = 2 \, \dm \, + \, V_{Coul}(R).
\eeq
In the above expression one reads the sum of $2 \, \dm$ and of the 
actual Coulomb static inter--quark potential $V_{Coul}(R)$. 
The logarithmic divergence connected 
to the renormalisation of the charge is contained in the potential. The 
latter actually defines a renormalised coupling in the so--called potential 
scheme according to 
\beq\label{potscheme}
V_{Coul}(R) \equiv - C_F \frac{\aV(R)}{R}.
\eeq
Note that till now we have made no use of the fact that we are going 
to compute all the quantities in the lattice scheme, apart from having 
explicitly written the linear divergence as $\dm$. By computing everything 
in lattice perturbation theory and by separating the linear from the 
logarithmic divergence one actually computes the coupling in the 
potential scheme ($\aV$) as an expansion in the lattice coupling ($\a0$), 
\ie one computes the matching between the couplings 
in the potential and the lattice scheme. In general at any given order such 
a matching is known once one knows the ratios between the $\L$ parameters 
and the differences between non--universal coefficients of 
the $\be$--functions $\{b_i, i \geq 2 \}$ in the two schemes. 
If we stop at the order we are interested in 
\beq\label{almatching}
\alpha_2(s \mu) = \alpha_1(\mu) + C_1(s) \, {\alpha_1(\mu)}^2 + 
C_2(s) \, {\alpha_1(\mu)}^3 + \, \ldots  
\eeq
where
\beqn
C_1(s) & = & - 2 b_0 \log s + 2 b_0 \log \frac{\L_2}{\L_1} \non \\
C_2(s) & = & {C_1(s)}^2- 2 b_1 \log s + 2 b_1 \log \frac{\L_2}{\L_1} + 
\frac{b_2^{(2)}-b_2^{(1)}}{b_0} \non
\eeqn
Now the point is that at this order the matchings of both the potential 
\cite{potms} and the lattice \cite{lattms} schemes to the $\MSbar$ scheme 
are known. Since the $\MSbar$ $\be$--function coefficients are known to 
three loops, one obtains from the above matchings 
both the ratios of $\L$ parameters and the $b_2$ coefficients for both the 
potential and the lattice schemes. This in turns means that the matching 
between the couplings in the potential and the lattice scheme is known to 
the order we need. This defines our strategy. From what we have just said 
it follows that in Eq.~(\ref{Vreq}) everything but $\dm$ is known, \ie
\beq\label{Vrmatch}
V(R) = 2 \, \dm \, - \frac{C_F}{R} \left( \a0 + c_1(R) \, \a0^2 + 
c_2(R) \, \a0^3 + \, \ldots \right)
\eeq
where
\beqn
c_1(R) & = & 2 b_0 \log R + 2 b_0 \log \frac{\L_V}{\L_0}, 
\;\;\;\;\;\;\; \log \frac{\L_V}{\L_0} = 4.4076 \non \\ 
c_2(R) & = & {c_1(R)}^2 + 2 b_1 \log R + 2 b_1 \log \frac{\L_V}{\L_0} + 
\frac{b_2^{(V)}-b_2^{(0)}}{b_0}, 
\;\;\;\;\;\;\; \frac{b_2^{(V)}-b_2^{(0)}}{b_0} = 8.5794 \non 
\eeqn
and
\beq
\dm  =  \sum_{n \geq 0} \, \overline{X_n} \, \a0^{n+1}, 
\eeq
the expression for $\aV$ having been taken from Eq.~(\ref{almatching}). 
The ``perfect'' strategy would then be as follows. Compute $V(R)$ in lattice 
perturbation theory and compare order by order the result to 
Eq.~(\ref{Vrmatch}): the coefficients in the expansion of $\dm$ are 
the only unknown quantities, that could be easily obtained. In practice 
we had to go through a number of approximations: we will discuss how they 
have to be taken under control and their impact on the final result 
estimated as a final error. First of all, we worked not in infinite volume, 
but on a lattice of fixed $32^4$ volume, as it is obviously requested by 
our computational tool (NSPT). We will further comment on this point later. 
Of course, this implies that we do not have access to the real $V(R)$, 
but only to its (finite volume) $V_T(R)$ approximations. Of course for 
small $R$ and $T$ and for finite lattice spacing the expression in 
Eq.~(\ref{Vrmatch}) is distorted by lattice artifacts. On top of all 
the above considerations one should also keep in mind that the results 
of NSPT are affected by statistical errors and their impact should be 
carefully assessed as well. In the end, what we did was to fit 
the $V_T(R)$ to the asymptotic expression for $V(R)$. The approximations 
$V_T(R)$ were in turn obtained from the computation of $W(R,T)$ for a wide 
range of $R$ and $T$. A first, obvious point to make is that taking 
lattice artifacts under control asks for having both $R$ and $T/R$ 
fairly large. The choice of a $32^4$ lattice was motivated by the 
compromise between these observations and the computing requirements of 
NSPT in terms of both memory and flops. Having said that, it is also 
important to point out from the very beginning that there are a couple 
of attitudes one can take with respect to the fits we had to perform. First, 
one can take the coefficients in the expansion of $\dm$ as the only unknown 
parameters, taking for all the 
other parameters entering Eq.~(\ref{Vrmatch}) their asymptotic, lattice 
artifacts free values. On the other side, one can study the stability 
of results by fitting the other constants in Eq.~(\ref{Vrmatch}) as well. 
This can give an idea of the extent to which our computations are 
good approximations of the infinite volume, lattice artifacts free results. 

\section{Results}
\subsection{$\dm$ to third order}
We computed $W(R,T)$ on a $32^4$ lattice for all $R,T \leq 16$. Since 
we needed to compute $\dm$ to order $\a0^3$, these expansions had to be 
computed to the same order. We refer people interested in these results 
themselves to a future review publication \cite{NSPT} in which they will 
be made available. In that context a detailed discussion of the statistical 
errors associated to NSPT results will be presented. For the purposes at 
hand here we only need to observe that a percent precision\footnote{We 
quote the worst precision, that is 
the one on order $\a0^3$.} of order $\leq 10^{-3}$ is easy to obtain in 
the range of $R$ and $T$ we were interested in. To do better than this 
would have been relatively easy, which thing we decided not to do in this 
exploratory study in the quenched approximation. Of course the statistical 
error is magnified by the extraction of the potential, since the $\log$ 
series are there. In the end, a percent precision only of order $10^{-2}$ 
was obtained for the potential at order $\a0^3$, again in the range of 
$R$ and $T$ we took into account. As far as the choice of the latter is 
concerned, we have already said that reducing lattice artifacts and 
keeping finite size effects under control were the major issues. Note that 
while the overall volume was kept fixed, both the $R/T$ and $R$ scales 
were relevant. It is trivial to observe that there is no finite size scaling 
that could be set up within these constraints. What we did was to constrain 
the values for $R$ and $T$ this way: for every choice of $T$ in the range 
$12 \leq T \leq 15$ the fits were done within an interval of $3 \div 4$ 
values of $R$ subjected to the constraints that {\em a)} the central 
value should be such that $R/T$ stayed around a value $R/T \sim 3$ 
and {\em b)} $R \geq 3$. It turns out that with such a procedure one is 
in a good position to perform the fit in which every parameter but 
the coefficients in the expansion of $\dm$ 
is taken as given by the asymptotic expressions. The results so obtained 
were regarded as approximations of the infinite volume, lattice artifacts 
free results. Of course 
on top of the statistical errors a systematic error has to be taken into 
account connected to the choice of the fitting interval, \ie of $R$ and $T$. 
Note that as approximations of the infinite volume results, the $V_T(R)$ 
should be loosely dependent on $T$. These can be tested also by studying 
the stability of results with respect to averaging as in $(V_T(R) + 
V_{T+1}(R))/2$. While this gives no new information at first and second 
order, it turns out that it lets you gain something at $\a0^3$ order, 
where the impact of statistical errors (which are bigger at this order) 
is a bit relieved by this averaging procedure. 
In order to gain some insight into the effectiveness of the procedure we 
have just described (that is, the one in which only $\dm$ is to be fitted), 
we first of all tried to reproduce the coefficients that were already known 
in the expansion of $\dm$. Expressing them in an expansion in the bare 
lattice coupling $\a0$, these coefficients are $\overline{X_0} = 2.1173$, 
$\overline{X_1} = 11.1520$. We got 
\beqn
\overline{X_0} = 2.118(2) \;\;\;\;\;\;\;\;\;\; 
\overline{X_1} = 11.14(3). \non
\eeqn
\begin{figure}[htb]
\begin{center}
\mbox{\epsfig{figure=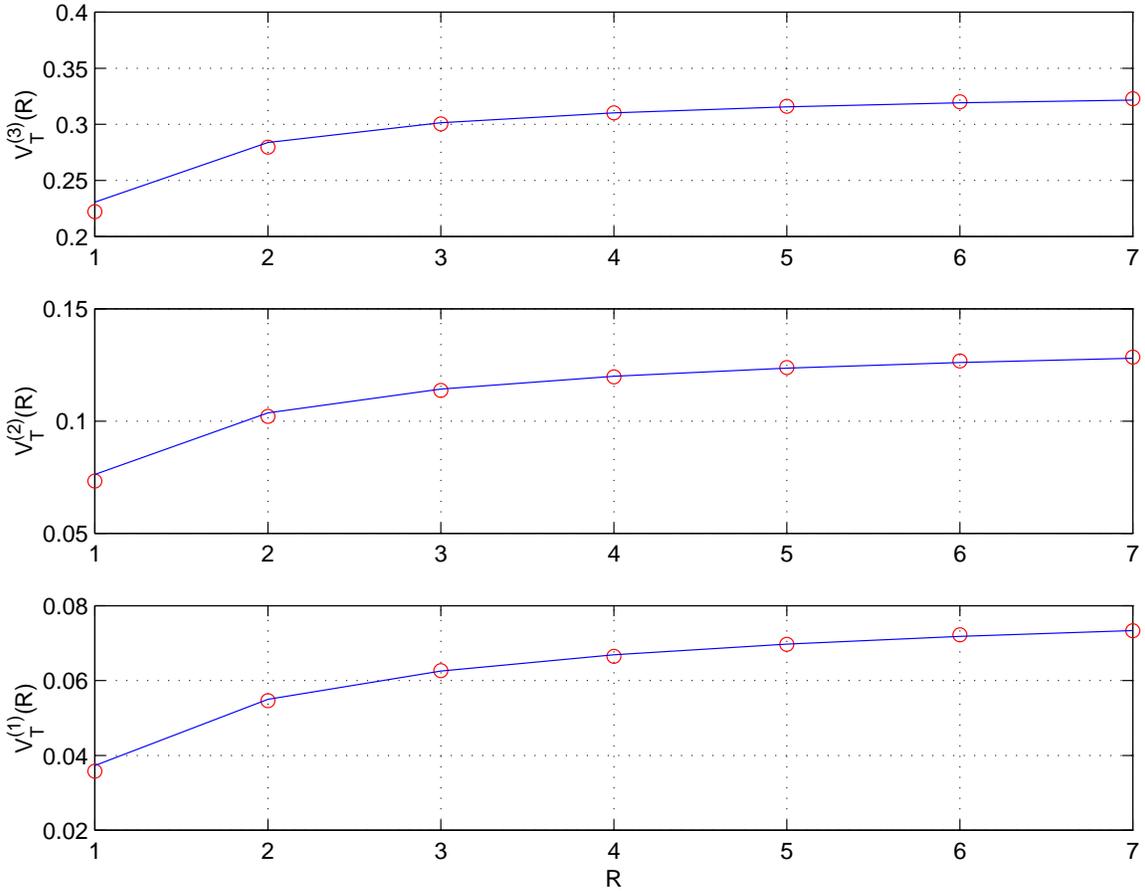,height=12cm}}
\caption
{A typical result of the fitting procedure described in the text. The 
three plots refer to first, second and third order of the potential. 
The solid lines are 
the appropriate orders as given by Eq.~(\ref{Vrmatch}) once $\dm$ has been 
fitted, while circles are the approximations $V_T(R)$. In this case $T=14$.}
\end{center}
\end{figure}T
The agreement with the analytical, infinite volume results is surprisingly 
good. The errors are in practice dominated by the systematic uncertainty 
connected to the choice of the fitting interval. Such an uncertainty was 
estimated to be given by the standard deviation pertaining to the sample of 
results of fitting procedures within the constraints we described a few 
lines above. For an extra peace of mind these systematic errors and the 
statistical ones were added not in quadrature. Before making further 
comments, one can proceed straight to 
repeat the same procedure at order $\a0^3$, which gives the 
result we are interested in as
\beqn
\overline{X_2} = 86.2(6). \non
\eeqn
This time an appreciable contribution is also coming from statistical 
errors in the computation of the loops and of the potential, even if 
the systematic uncertainty connected to the choice of the fitting interval 
nevertheless stays as the major one. Again, these different errors were 
added not in quadrature. In the figure we plot an example of the results 
one gets by this procedure: by fitting values for the coefficients in the 
expansion of $\dm$ one obtains that order 
by order the approximation of $V(R)$ falls on the curve given by 
Eq.~(\ref{Vrmatch}). One should keep in mind that the effectiveness of 
the procedure is actually magnified by the figure. This in particular 
true for the highest order, for which the fitted result is a huge shift 
with respect to the final order of magnitude of plotted data. The 
sensitivity with respect to such a huge shift can be easily underestimated 
if one only looks at the figure. Note by the way that the huge shift 
gives a good idea of the impact of the renormalon on the expansion. 
Now the question comes whether the error bars obtained in the first 
procedure we described are to be fully trusted. As we have already pointed 
out the effectiveness in reproducing the known coefficients within fairly 
tiny errors is impressive. As it is obvious, this is nevertheless only 
known {\em a posteriori} and in this sense it is not a proof, but only a 
hint at the reliability of the method. As stressed, our first procedure 
relied on the assumption that the $V_T(R)$ were good approximations of the 
infinite volume, lattice artifacts free $V(R)$, the only sources of errors 
being the statistical 
ones and the systematic effects of the choice of the fitting intervals. 
As we have already suggested, a second attitude can be taken with 
respect to fitting the $V_T(R)$ to the expression given by 
Eq.~(\ref{Vrmatch}). This does not assume the other coefficients 
entering the formulas as given by their asymptotic, lattice artifacts 
free values. This means at leading order to fit also the 
coefficient $C_F$, at second order also the ratio of the $\L$ parameters and 
at third order the difference in the $\be$--function coefficients $b_2$ 
as well. This is an indirect test of how asymptotic our approximations are. 
One should anyway always keep in mind that this is not a good starting 
point for a proper finite size analysis, since in all our procedure there is 
a finite fixed volume and a couple of scales around. We will anyway assume 
the typical variation of coefficients with respect to this procedure 
as a sensible bound on finite size and lattice artifacts effects. 
The constraints we imposed 
before on the choice of the fitting intervals had to be somehow released in 
order to be able to fit more than one parameter. Anyway, no more than two 
parameters were fitted simultaneously (that is, at every order the 
relevant coefficient in the expansion of $\dm$ plus another 
constant). This was obtained by letting every order inherit fixed values 
for other parameters. For example: the second order fit was in charge of 
fitting $\overline{X_1}$ and the ratio of $\L$ parameters, inheriting 
the value for $C_F$ either from first order fit or from the infinite volume 
result (of course also variations with respect to these variants 
were taken into account). 
Depending again on the choice of fitting intervals, we found a rough 
bound of $10 \div 15\%$ on the variations of all the parameters entering 
the Coulomb potential, including $\frac{b_2^{(V)}-b_2^{(0)}}{b_0}$. 
Ironically, the latter used to be for us a 
first source of confidence in our results. Given our ignorance of reference 
\cite{potms}, at the beginning we fitted this coefficient together with 
the third order $\dm$, assuming for the other parameters their asymptotic 
values. The reasonable agreement of the result of the fit with \cite{potms} 
turned out to be a reasonable hint at the reliability of our result for 
$\overline{X_2}$. As far as the values of the coefficients of the expansion 
of $\dm$ we found instead a rough bound of $1\%$. Note how this bound 
appears surprisingly huge with respect to the deviation of our results 
from the exact ones for first and second order. As we have already 
observed, this is something which can be anyway verified {\em a 
posteriori}. A possible step forward in the estimate of finite size 
effects could be to repeat the whole procedure on a different volume, 
which thing we propose to do in the future unquenched study. 
At the moment, taking into account also the last considerations, for an 
extra piece of mind we assume as our final result 
\beqn
\overline{X_2} = 86.2(0.6)(1.0). \non
\eeqn
Our result is quite apart from the one \cite{mack} presented at the Lattice 
99 conference by the Fermilab group, suggesting a fairly big impact of 
systematics. We point out that the computational tools are as a matter of 
fact quite different. What we plan to do in the future (for the unquenched 
computation) is to follow also their approach based on Polyakov instead 
of Wilson loops, which thing allows a fine finite size analysis. 

\subsection{The impact of the result}
It is now the proper time to address the question of the impact of our 
result. As a matter of fact the result presented in this paper has already 
been known for some time to people involved in the field. Lubicz's review 
on quark masses at the Lattice 2000 conference \cite{Vitt} has already 
referred to the impact of the knowledge of $X_2$ in the determination of 
the $b$--quark mass, at least in the quenched approximation. Basically, 
things go as follows \cite{Vice} (this is what one would refer to as 
a Next to Next to Next to Leading Order - NNNLO - quenched calculation of 
the $b$--quark mass). The lines along which the computation can go 
have already been widely presented in the Introduction, the final 
formula being Eq.~(\ref{matching}). In order to understand what one 
achieves with the result for $X_2$ a good point is the comparison with 
the lower order (NNLO) determination\footnote{The reader can get all the 
details for lower orders results in references \cite{CGMS95,MS98,GGMR00}.}. 
Of course all the following considerations about the impact of NNNLO hold 
for the quenched case, given the limitation of our result. The residual 
error due to unknown perturbative corrections in Eq.~(\ref{matching}) is 
halved with respect to NNLO and the new order improves the convergence of 
the series. Given that the determination of the quark mass can be got from 
results obtained at various values of $\be$, a remarkable point 
is that the dependence of the 
final result for the quark mass on the lattice spacing is smaller than in 
NNLO and very weak. In the end, it turns out that the overall error on 
the mass is of order $1 \div 2\%$, the contributions being equally split 
between perturbative and non--perturbative ones (this final error was at 
NNLO dominated by the ignorance of $X_2$).

\section{Conclusions}
We computed the perturbative expansion of the residual mass term in lattice 
Heavy Quark Effective Theory to order $\a0^3$. This is an important 
building block in a renormalon safe determination of the $b$--quark mass 
from lattice simulations. Errors appear to be under a fairly good control 
and the impact of the result on the (unquenched) determination of the 
quark mass is quite important. Nevertheless, the result we presented is 
still a partial one: going for unquenching is compelling. The latter is 
our proposed next step forward. We will do that trying to still improve on 
systematics. 

\section*{Acknowledgments}
\par\noindent
We thank G. Burgio and M. Pepe for having collaborated 
with us at an early stage of a work which eventually also merged in 
this paper. We are extremely grateful to V. Gimenez and V. Lubicz 
for illuminating discussions on the subject and to G. Martinelli and 
C.T. Sachrajda not only for what we learned from them on the subject, but 
also for their constant interest and encouragement. We also thank P. 
Mackenzie and the Fermilab group for having shared with us their 
understanding of the subject. 
F.~D.R. acknowledges support from both Italian MURST 
under contract 9902C68583 and from I.N.F.N. under {\sl i.s. PR11}. L.~S. 
acknowledges support from EU Grant EBR-FMRX-CT97-0122.

%=====================================================================


\begin{thebibliography}{99}
\bibitem{HQET}
	M. Neubert, \Journal{\PR}{245}{259}{1994}
\bibitem{FNL92}
	A. F. Falk, M. Neubert, M. E. Luke, \Journal{\NPB}{388}{363}{1992}.
\bibitem{BB94}
	M. Beneke, V.M. Braun, \Journal{\NPB}{426}{301}{1994}.
\bibitem{MMS92}
	L. Maiani, G. Martinelli and C. T. Sachrajda,
	\Journal{\NPB}{368}{281}{1992}.
\bibitem{MS95}
	G. Martinelli and C. T. Sachrajda, \Journal{\PLB}{354}{423}{1995}.
\bibitem{CGMS95}
	M. Crisafulli, V. Gim\'enez, G. Martinelli and
	C. T. Sachrajda, \Journal{\NPB}{457}{594}{1995}.
\bibitem{GGMR00}
	V. Gim\'enez, L. Giusti, G. Martinelli and
	F. Rapuano, \Journal{\JHEP}{03}{18}{2000}.
\bibitem{CSR99}
	K.G. Chetyrkin and M. Steinhauser, \Journal{\PRL}{83}{4001}{1999}; 
	K.G. Chetyrkin and A. Retey, \Journal{\NPB}{583}{3}{2000}.
\bibitem{MR99}
	K. Melnikov and T. van Ritbergen, \Journal{\PLB}{482}{99}{2000}.
\bibitem{NSPT}
        F. Di Renzo, G. Marchesini, P. Marenzoni and E. Onofri,
	\Journal{\NPB}{426}{675}{1994}. For a recent update see also 
	F. Di Renzo, L. Scorzato, \Journal{\NPBPS}{83}{822}{2000} and 
	F. Di Renzo and L. Scorzato, {\em ``Understanding Numerical Stochastic 
	Perturbation Theory: a status report for Lattice Gauge Theories''}, 
	in preparation.
\bibitem{MS98}
	G. Martinelli and C. T. Sachrajda, \Journal{\NPB}{559}{429}{1999}.
\bibitem{lat99}
	G. Burgio, F. Di Renzo, M. Pepe, L. Scorzato, 
	\Journal{\NPBPS}{83}{935}{2000}.
\bibitem{mack}
	G.P. Lepage, P.B. Mackenzie, N.H. Shakespeare, H.D. Trottier, 
	\Journal{\NPBPS}{83}{866}{2000}.
\bibitem{DV80}
	V. S. Dotsenko, S. N. Vergeles. \Journal{\NPB}{169}{527}{1980}
\bibitem{HK}
	U. Heller, F. Karsh, \Journal{\NPB}{251}{254}{1985}. 
\bibitem{potms}
	Y. Schr\"{o}eder, \Journal{\PLB}{447}{321}{1999}.
\bibitem{lattms}
	M. Luescher, P. Weisz, \Journal{\NPB}{452}{234}{1995}. 
	B. Alles, A. Feo, H. Panagopoulos, \Journal{\NPB}{491}{498}{1997}. 
\bibitem{Vitt}
	V. Lubicz, hep-lat/0012003, to be published in the Proceedings of 
	the Lattice 2000 conference. See also S. Collins, hep-lat/0009040, 
	to be published in the Proceedings of the Confinement IV conference. 
\bibitem{Vice}
	V. Gim\'enez and G. Martinelli, private communication. 
\end{thebibliography}
\end{document}